# Machine Culture

Levin Brinkmann,*[†][1] Fabian Baumann,[†][1] Jean-François Bonnefon,[†][2] Maxime Derex,[†][2,4] Thomas F. Müller,[†][1] Anne-Marie Nussberger,[†][1] Agnieszka Czaplicka,[1] Alberto Acerbi,[3] Thomas L. Griffiths,[5] Joseph Henrich,[6] Joel Z. Leibo,[7] Richard McElreath,[8] Pierre-Yves Oudeyer,[9] Jonathan Stray,[10] Iyad Rahwan*[†][1]

* Correspondence: brinkmann@mpib-berlin.mpg.de; rahwan@mpib-berlin.mpg.de
[†] Equal contributions

[1] Max Planck Institute for Human Development, Center for Humans & Machines, Lentzeallee 94, Berlin 14195, Germany
[2] Toulouse School of Economics, Esplanade de l'Université, 31080 Toulouse Cedex 06, France
[3] Department of Sociology and Social Research, University of Trento, Italy
[4] Institute for Advanced Study in Toulouse, Esplanade de l'Université, 31080 Toulouse Cedex 06, France
[5] Department of Psychology & Department of Computer Science, Princeton University, Princeton, NJ 08540
[6] Department of Human Evolutionary Biology, Harvard University, 11 Divinity Ave, Cambridge, MA 02138, USA
[7] DeepMind Technologies Ltd, London, United Kingdom
[8] Max Planck Institute for Evolutionary Anthropology, Leipzig, Germany
[9] Inria, Flowers team, Université de Bordeaux, France
[10] Center for Human-Compatible Artificial Intelligence, University of California, Berkeley, USA

**Abstract:**
The ability of humans to create and disseminate culture is often credited as the single most important factor of our success as a species. In this Perspective, we explore the notion of 'machine culture,' culture mediated or generated by machines. We argue that intelligent machines simultaneously transform the cultural evolutionary processes of variation, transmission, and selection. Recommender algorithms are altering social learning dynamics. Chatbots are forming a new mode of cultural transmission, serving as cultural models. Furthermore, intelligent machines are evolving as contributors in generating cultural traits–from game strategies and visual art to scientific results. We provide a conceptual framework for studying the present and anticipated future impact of machines on cultural evolution, and present a research agenda for the study of machine culture.

# Introduction

The ability of humans to create and disseminate culture is considered the single most important factor in our species' dominance on earth [1]. The evolution of human culture has been the subject of extensive study in all of the behavioral sciences, including anthropology [1], psychology [2], cognitive science [3], biology [4], linguistics [5,6], archaeology [7], sociology [8] and economics [9] (Box 1).

Cultural evolution exhibits key Darwinian properties. Culture is shown to exhibit variation, transmission, and selection, and evolves through the selective retention of cultural traits, as well as nonselective processes such as drift [10]. Major shifts in any of these three Darwinian properties can greatly impact cultural evolution. For instance, between 1300 and 1600, European culture experienced successive major shifts due to increased exposure to Chinese technology such as gunpowder, which changed the nature of warfare (variation)[11]; Gutenberg's invention of the printing press (transmission)[12]; and renewed interest in Classical ideas and values, such as Classical ideals of artistic expression, during the Renaissance (selection)[13]. When such substantial changes occur, they induce rapid and major impact on culture.

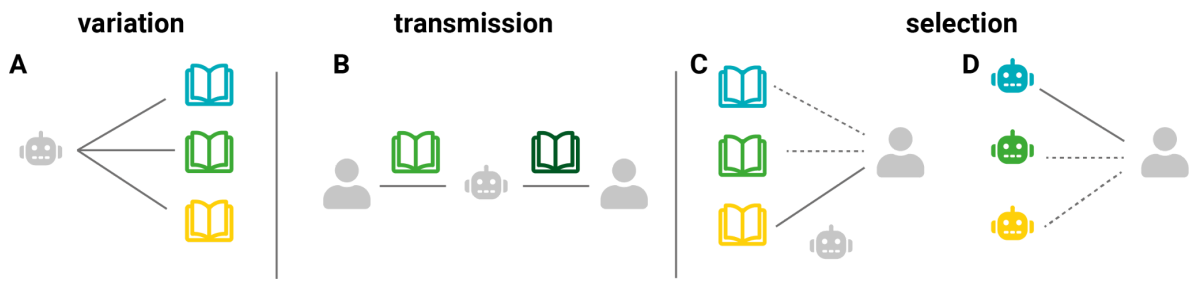

**Figure 1: Examples of machine culture. A**. Generation of novel cultural artifacts through machines. **B**. Machine transmits and potentially mutates cultural artifacts. **C**. Machine selects between different cultural artifacts. **D**. Human selects among diverse machines.

While new technologies have always affected the course of cultural evolution, in this article we argue that intelligent machines will exert a transformative influence on cultural evolution through their impact on all three Darwinian properties of culture: variation, transmission, and selection (Fig. 1). This process began in the early days of the Internet with machine-based content ranking by search engines and social media feed algorithms influencing what information people get from others. More recently, generative algorithms have begun participating in the creation of cultural traits themselves. We are not only observing a transformation of human culture but also its evolution into machine culture—culture mediated or generated by machines. This article aims to provide researchers across disciplines with a primer and a roadmap for navigating this monumental shift. As the impact of an increasingly digital society on cultural evolution has been explored elsewhere [14], we specifically focus on the current and potential impact of intelligent machines on cultural evolution. For the purposes of this article, we use the terms "intelligent machines" and "artificial intelligence (AI) systems" interchangeably, with AI referring to the science and technology that allow machines to perform tasks that typically require human intelligence such as perceiving the environment, planning and executing actions, and adapting by learning from data or experience [15,16].

## Examples of machine-mediated cultural evolution

We begin by presenting empirical evidence of machine cultural evolution, setting the stage for a detailed exploration through a framework that discusses instances where machines mediate or generate cultural traits from a cultural evolutionary perspective. Through four pivotal examples, we illustrate the diverse ways intelligent machines are transforming cultural evolutionary dynamics. Generative machines, such as text-to-image algorithms, are contributing to the variety of cultural traits. Models drawing upon reinforcement learning are pushing humans onto novel ground, for instance in the ancient game of Go and beyond. Large language models (LLMs) are facilitating the transmission of cultural knowledge and redefining the value of human intellectual skills. Meanwhile, transmission pathways are rewired by recommender systems selecting what and from whom humans learn. At first glance, the examples provided might seem to pertain to vastly different technological areas, and to translate into a collection of unrelated effects. However, even now, machines are beginning to integrate features from a range of the outlined technologies—reinforcement learning, for instance, is enhancing generative AI. Furthermore, these technologies are operating on multiple levels; generative AI not only generates novel ideas but also offers recommendations for their refinement.

> **Box 1: Glossary**
>
> **Culture**: information capable of affecting individuals' behaviors that are acquired from other individuals via social transmission
> **Cultural Evolution**: the change of cultural information over time, the key properties for an evolutionary process are variation, transmission, and selection
> **Social Learning**: learning that is influenced by the observation of another individual or their products.
> **AI**: the science and technology enabling machines to perform tasks that typically require human intelligence such as perceiving the environment, planning and executing actions, and adapting by learning from data or experience.
> **Machines**: intelligent machines. Used interchangeably with AI systems, thus referring to machines that may possess capabilities to perceive the environment, plan and execute actions, and adapt by learning from data or experience.
> **Variation**: the existence of different cultural traits within a population. It represents the raw material on which other processes, like selection and transmission, operate. Humans and machines add to existing cultural variation through random and guided exploration, as well as recombination of existing cultural traits.
> **Transmission**: the process by which cultural information, such as knowledge, behaviors, traditions, or practices, is passed from one individual to another through social learning mechanisms such as observation or teaching.
> **Selection**: the process by which certain cultural traits, practices, or ideas become more or less prevalent within a population over time due to differential adoption.

*Cultural recombination through generative AI*

Generative AI has seen two major waves of innovation in recent years. The inception of Generative Adversarial Networks (GANs) by Goodfellow et al. in 2014 enabled the algorithmic generation of high-fidelity images [17]. GANs offered capabilities beyond the generation of lifelike images—they also have the ability to blend or interpolate—giving birth to novel creations such as fantasy lifeforms [18]. Subsequent advancement in 2022 saw the advent of diffusion-based text-to-image generative AI systems such as DALL·E, Midjourney, and Stable Diffusion. These models substantially enhanced the recombination power of these early models by generating high-resolution images conditioned on text descriptions [19–21]. The original DALL·E, although now surpassed by other models in terms of image quality, demonstrated such recombination capabilities impressively [19]. For instance, when prompted to produce 'an armchair in the shape of an avocado', it creatively recombined these two distinct concepts (see Fig. 2).

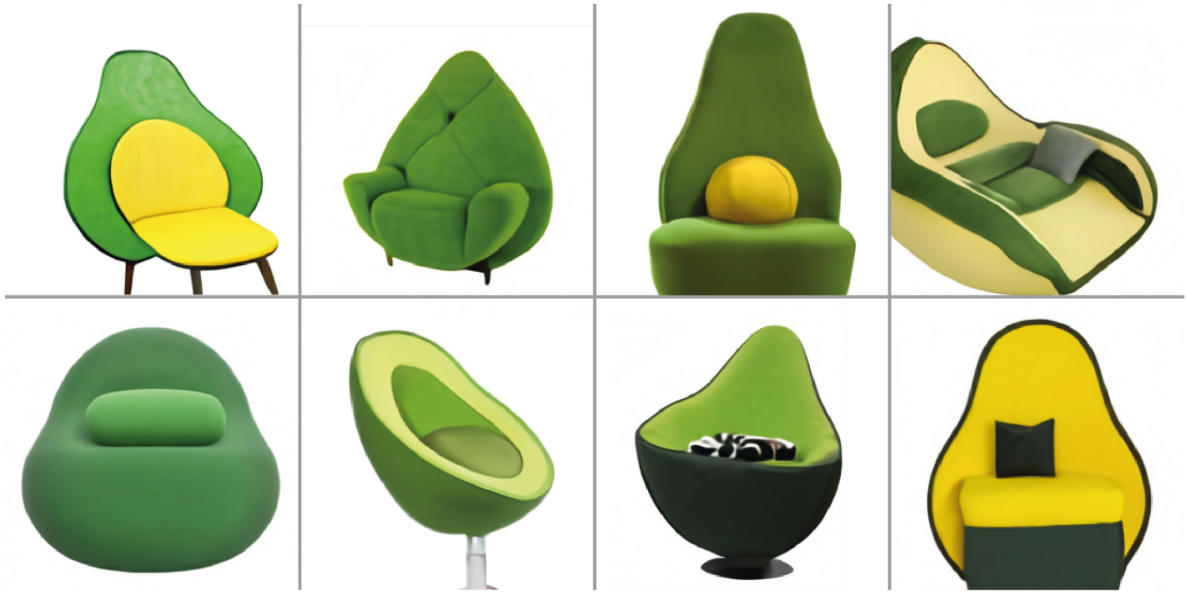

**Figure 2: Recombination of Visual Concepts**. The avocado chair, synthesized by OpenAI's text-to-image generative AI, DALL·E, exemplifies the early stages of algorithmic cultural recombination. By seamlessly combining previously learned concepts—avocados and chairs in this case—the model showcases the ability to create coherent integrations of disparate elements and demonstrates novelty through recombination.

These models can thus increase cultural variation by helping humans to produce new and relevant recombinations, which are sometimes recognized as works of art, sold at prestigious auction houses [22]. While recombination oftentimes forms the foundation of human creativity [23], it is still debated how, or even if, machines can generate relevant content beyond the boundaries of human culture. Even simple latent representations can disentangle the semantic meaning of linguistic concepts [24]. Similarly, text-to-image models use language as a cognitive tool to disentangle and consequently recombine visual concepts [25]. However, these models build upon concepts harvested from human culture. As such, text-to-image models may be limited to the concepts defined and demonstrated by humans in the underlying dataset. However, generative AI models can produce novel "future art" by forecasting future art movements and by deliberately avoiding classification into established artistic movements—such creations have been evaluated as more creative than prestigious contemporary works by human artists [26]. When applied to engineering, similar methods can lead to the discovery of designs that are both novel and superior [27].

Generative AI exemplifies the dual nature of machines as both cultural artifacts and creators thereof. On the one hand, generative algorithms are increasingly presented and, to some extent, recognized as authors of art [22]. Simultaneously, machines are subjected to cultural processes of comparison, distribution, modification, and eventual abandonment.

*Cultural innovation through reinforcement learning*

In 2016, AlphaGo defeated Lee Sedol, the world champion Go player, with a series of four victories over five games. Remarkably, AlphaGo managed to surprise Sedol with distinctively nonhuman gameplay. In particular, move 37 in the second game was considered extremely unconventional, estimated by AlphaGo itself to have a 1 in 10,000 chance of being made by a human [28] (see Fig. 3.A). AlphaGo's unconventional gameplay likely originated in its self-play training: While its training started with reproducing human gameplay from a large database, AlphaGo also trained through self-play, selecting promising but uncertain moves and evaluating their success against itself [29]. Thereby, it iteratively improved and developed novel game strategies. It was one of these innovative, self-trained strategies that took Lee Sedol by surprise. As it turned out, it was not even necessary for the model to start from learning human gameplay at all. The successor of AlphaGo, named AlphaGo Zero, ignored all of the accumulated Go knowledge of humankind [29]. Starting from a blank slate, it not only rediscovered human Go strategies but also developed strategies that surpassed those of its human creators.

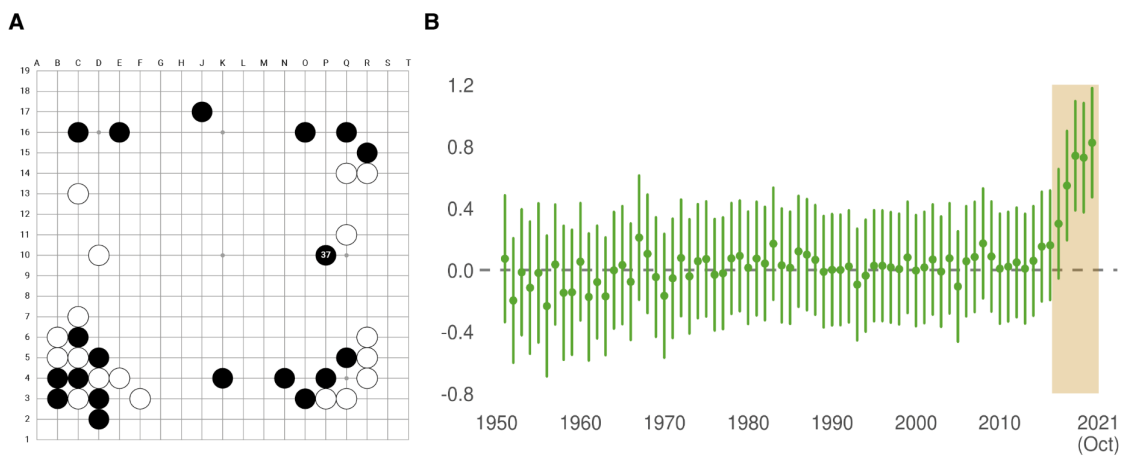

**Figure 3: Go play before and after the introduction of AlphaGo.** **A** AlphaGo, in its match against Go world champion Lee Sedol, made a highly unusual and strategic 37th move by placing its stone further from the edge, towards the center of the board, deviating from the traditional strategy of securing territory along the periphery during the early stages of the game. With this unconventional move, AlphaGo not only broke with centuries-old Go traditions but also paved the way for its ultimate victory in the match. **B** (reproduction based on [30]) Decision quality of professional Go players as evaluated by an algorithm performing at superhuman level. Decision quality significantly increased after Lee Sedol was beaten by AlphaGo on March 15, 2016 (shaded area).

The innovations generated by AlphaGo and AlphaGo Zero soon entered human culture, as shown by research comparing human gameplay before and after the algorithms' introduction [30]. The decision quality, as measured by an open-source variant of AlphaGo Zero, showed very little improvement in human gameplay from 1950 to 2016, followed by a sudden improvement after the introduction of AlphaGo in March 2016 [31,32] (see Fig. 3.B). However, this improvement wasn't solely due to humans adopting strategies developed by AlphaGo. It also reflected an unexpected shift, wherein humans started developing moves that were qualitatively distinct both from previous human moves and from the novel moves introduced by AlphaGo. In summary, AlphaGo served as an early, quantifiable exemplar of machine culture, generating novel cultural variations through genuine, nonhuman innovation. This was followed by a major transition into an even broader range of traits as the result of humans building on the previous discoveries made by machines. As the methods underpinning AlphaGo have been generalized to other games and extended to scientific problems, we anticipate a continued infusion of machine-generated discoveries across diverse domains of human culture [33,34].

*Language models transmit and revalue cultural knowledge*

The release of ChatGPT, a widely accessible LLM, has revolutionized how we interact with machines, using it to learn, brainstorm, and refine ideas. Trained on extensive human text data, both historical and contemporary, LLMs act as models of human culture [25], facilitating cultural transmission across individuals and generations. In due course, students began requesting LLMs to complete their homework [35]; knowledge workers used LLMs (at their own peril) to automatically extract and summarize required content [36]; and software developers widely adopted LLMs as powerful code-writing assistants [37]. LLMs not only serve as content creators but also, for better or worse, act as reservoirs of knowledge and providers of learning opportunities. Thereby, ChatGPT exemplifies social learning from machines.

As the capabilities and usages of LLMs continue to develop, the value of certain human skills will shift. Some skills may lose value quickly, especially in language-related and cognitively demanding occupations such as translation, copywriting, or proofreading [38]. Occupations with more creative uses of language may follow, given that LLMs may soon surpass the creativity of humans as measured by standardized tests [39]. Even though not all occupations will be affected in the short-term, a recent survey projects that around 20% of the workforce will experience LLMs impacting at least half of their tasks [38]. It is important to note, however, that this projection is an early estimate and as such inherently uncertain in its accuracy and reliability. Meanwhile, other skills may gain in value — for instance, skills that allow efficient collaboration with LLMs, such as prompt engineering (i.e., the skilful writing of instructions to get LLMs to do what we want) [40]. Consequently, human creativity is experiencing a remarkable shift [41]. It is not merely focused on generating final outputs but is increasingly evolving towards interactions with machines, progressing from explicit prompts to more natural conversations [42]. Workers who invest in related skills may outperform workers who do not, potentially accelerating the adoption of LLMs and further increasing the value of knowing how to collaborate with them.

*Cultural rewiring through recommender systems*

The digital age is so data-rich that it has become increasingly hard for humans to navigate available information. In this abundance, recommender systems that manage and filter information have a silent but increasingly important role that is easily overlooked given how seamlessly these systems have integrated into our everyday digital lives. These systems select and prioritize items based on a variety of explicit and implicit features, including personal interests, control settings, past user behavior, and the behavior of similar users. While recommender systems do not add variation in cultural traits, they demonstrably impact the selective retention and transmission of cultural traits.

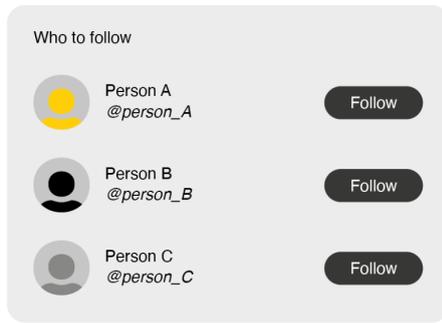
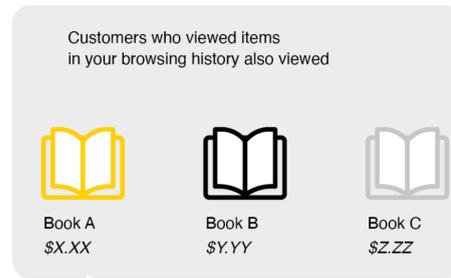
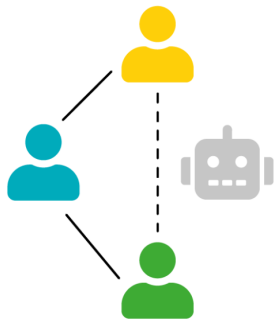
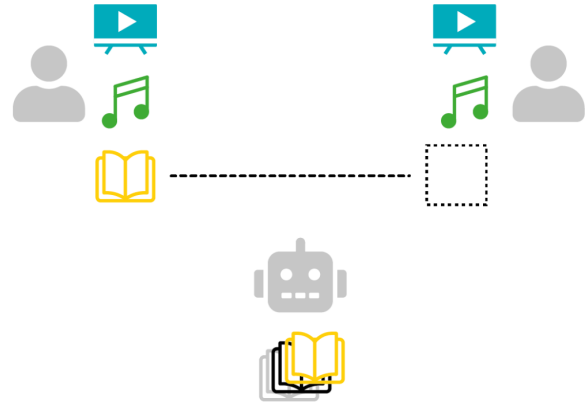

**Figure 4: Exemplary instances of cultural rewiring.** Top panels represent the user interfaces, bottom panels provide a schematic depiction of the underlying mechanism. **A** Friendship recommendation on an online platform (e.g., X or LinkedIn). A new social tie (person A) is recommended to the green user. **B** Collaborative filtering for item recommendations. The recommender system suggests a book to the right user based on the correlations in co-purchases of other items (movie and music album) by the left user.

By promoting new social ties—such as suggesting who to follow on X (formerly Twitter), to date on Tinder, or to work with on LinkedIn—recommender systems alter the people to whom we are exposed to, ultimately changing the structure of our social networks and hence pathways of cultural transmission [43] (see Fig. 4.A). But recommender systems can also bypass network structures, exerting an even more direct impact on which cultural content or products we are exposed to. For example, e-commerce websites and streaming platforms deploy recommender systems to steer customers through the expansive array of available products based on content and collaborative filtering. Content filtering matches information about a customer's consumption history with the attributes of all available products to make suggestions about related purchases: a customer who has recently purchased running shoes may receive suggestions for additional running equipment, matched to the shoes in price and design [44]. Meanwhile, collaborative filtering [45] makes recommendations based on less obvious patterns of correlations in users' profiles: if users A and B overlap in their previously consumed movies and music, the recommender system might suggest content of a completely different kind, for example, recommending to user A a book that user B liked (see Fig. 4.B). In sum, recommender systems influence cultural evolution by rewiring our social networks and modifying information flows such that they can substantially influence the dynamics of cultural markets [46–48].

# A framework for machine-mediated cultural evolution

Building on these exemplary instances of how machine technologies may impact cultural evolution, we will now map out a systematic framework for studying the potential of machines to shape cultural evolutionary processes.

Culture has been defined as information capable of affecting individuals' behaviors that is acquired from other individuals via social transmission [49]. Consequently, the science of cultural evolution examines the change of cultural information over time [50,51]. Cultural information is represented by individual cultural traits, which can exist as cognitive representations or be expressed in behaviors or artifacts [52]. Culture evolves according to a similar process by which species change, that is, through the selective retention of cultural traits and through other nonselective processes, such as drift [10]. While there are ongoing discussions on how far the analogy between cultural and genetic evolution should be pushed [53,54], there is general agreement that culture exhibits the key properties of evolutionary systems that are Variation, Transmission, and Selection. These properties are not necessarily the result of distinct processes of human cognition and behavior, yet they offer a useful framework to analyze the multifaceted ways machines can influence cultural evolution (see Table 1 for a summary).

Variation refers to the existence of different cultural traits within a population. Transmission involves the spread of cultural information from one individual to another through social learning mechanisms, including observation and teaching. During this transmission, information losses often occur, affecting the preservation of cultural traits. Selection occurs when certain cultural traits are more likely to be adopted by individuals due to factors such as their usefulness, popularity, or compatibility with existing cultural practices. Crucially, the prevalence of specific traits over time can be influenced by both their selection and variations in transmission fidelity, contingent on the traits' characteristics. We anticipate that machine technologies will affect each of these three properties(see Fig. 4), which are likely to have transformative impacts on cultural evolution.

**Table 1.** Tentative conjectures about ways in which machines might shape the processes of cultural evolution.

| | **Machine Capability** | **Possible Impact on Culture** |
|---|---|---|
| **Variation** | Learning at unprecedented scale and speed (e.g., reinforcement learning) | Emergence of solutions culturally alien to humans |
| | Superhuman model complexity | Generation of solutions inconceivable by collective human intelligence and human cultural evolution |
| | Incomparably broad and deep knowledge base | Creation of novel recombinations beyond the human horizon |
| | (Semi-)autonomous creativity (e.g., image generation models) | Facilitation or crowding out human participation in creative exploration |
| **Transmission** | Exposing and/or preserving documented cultural knowledge (e.g., LLM chatbots) | Enhanced retrieval and increased transmission fidelity of documented cultural knowledge |
| | Reproduction of human biases | Amplification or mitigation of existing biases; potential for increased cultural erosion |
| | Accelerated processing of empirical evidence | Facilitation of transmission of less compressed knowledge |
| | Leveraging the unique cognitive capabilities of both humans and machines | Expanding the collective capacity to maintain diverse cultural artifacts |
| **Selection** | Recommendation of social ties (e.g., link recommendation algorithm) | Shaping social networks, potentially inhibiting or enhancing serendipitous encounters |
| | Curating content (e.g., ranking algorithms, collaborative filtering) | Indirectly shaping social networks via content exposure |
| | | Shaping incentives for human content creators (e.g., clickbait) |
| | Adapting to human feedback (e.g., RLHF) | Alignment of machines to human goals, potentially leading to unintended consequences (e.g., spread of highly believable myths) |
| | Machine learning from machine-generated content | Selection of machine-generated content as main driver of cultural evolution |
| | Adaptability of AI models to market/societal demands | Proliferation of appealing apps with varying alignment to human welfare |
| | Competing with humans in cultural production (e.g., poetry) | Specialization of humans and machines in distinct niches |

*Variation*

Variation refers to the presence of diverse cultural traits within a population. Humans and machines contribute to this existing cultural variation through both random and guided exploration, as well as the recombination of existing cultural traits. However, machines, leveraging their unique capacities, can produce traits distinct from those produced by humans, thus potentially steering culture toward new paths.

Machines have the ability to learn individually at an unparalleled scale, enabling the discovery of novel cultural traits through extensive exploration. Thorough exploration is not unique to machines. Thomas Edison, for example, famously tested over 6,000 materials to find the most suitable filament for his incandescent light bulb [55]. However, the computational speed of humans imposes time constraints that limit their exploration compared to machines [56]. For instance, AlphaGo Zero needed only three days to play 4.9 million games against itself, achieving a superhuman level of proficiency [29]. No human can amass that volume of experience individually; hence, we tend to rely heavily on preexisting, culturally evolved solutions that are socially learned [57,58]. This approach narrows the scope of innovation to the cultural context of previous generations. AI systems like AlphaGo use reinforcement learning—based on iterative rounds of trial and error—and hence can transcend this cultural path dependency by starting from a blank slate and discovering innovative solutions through exploration alone [59]. As a result, AI systems have the potential to generate culturally alien traits. In this respect, AlphaGo is in stark contrast to language models, which primarily learn to reproduce human cultural output.

Machines and humans employ guiding models—policies—that aid in efficiently exploring complex solution spaces, thereby enabling the discovery of solutions that would be unobtainable through random exploration alone. For instance, Mesoamerican skywatchers, via an early example of human astronomic modeling, developed a policy for optimal seed planting times, maximizing agricultural productivity [60]. Similar to humans, algorithms like AlphaGo utilize complex policies to guide their search. The general approach to guided search is similar for both humans and machines: experience is used to update a predictive model of the world, which then informs actions that generate new experiences. Machines no longer require humans to explicitly formulate these guiding models. Instead, general-purpose neural networks are universal approximators, allowing functional relationships to be learned [61,62]. This allows for unprecedented model complexity and consequently pushes the boundary of attainable solutions [63]. Surpassing the capabilities of any human-conceived model, the neural network-based algorithm AlphaFold has achieved remarkable precision in modeling intramolecular relations, enabling it to tackle one of molecular biology's greatest challenges: predicting a protein's three-dimensional structure based solely on its DNA sequence [64]. For centuries, protein structure prediction has drawn substantial intellectual contributions from scientists, even harnessing the collective intelligence of tens of thousands of citizen scientists [65]. Nevertheless, AlphaFold's complex model of intramolecular relations, manifested in the form of a deep neural network, proved superior, eclipsing these extensive human efforts with remarkable proficiency.

Many contemporary algorithms, such as LLMs, have been trained on a cultural repertoire of unprecedented scale [66]. For instance, some LLMs are trained on hundreds of billions of words [67]. While size does not guarantee diversity [68], the degree to which machines can learn from human culture repertoires far exceeds what can be achieved by a single human being, allowing, for instance, individual language models to cover a broad spectrum of languages comprehensively [69]. By combining knowledge from disjointed communities—spanning cultural and geographic contexts, language barriers, or scientific disciplines—intelligent machines can produce novel recombinations that may be beyond humanity's reach. Irrespective of whether LLMs truly exhibit understanding [70], the mere recombination of cultural traits can lead to innovation and fuel cultural evolution [68,71]. Indeed, recombination is considered central to generating novel cultural traits [23,68,72–74]. Another way in which intelligent machines may increase the variety of cultural traits is by building on human knowledge while at the same time identifying and deliberately avoiding common human pathways. For instance, in the realm of scientific discoveries, intelligent machines may be designed to uncover scientifically plausible and promising "alien" hypotheses that would normally fall outside the focal point of contemporary scientific communities [75]. This does not imply, however, that intelligent machines are without their own limitations or blind spots, as we will discuss later.

But machines may also augment human exploration. Conditional generative AI, like DALL-E, allows users to steer the generative process through text prompts, enabling experimentation with diverse concepts and visualizations without the need for advanced artistic skills [19]. Throughout human history, the effective size of the population that contributed to the creation and maintenance of cultural knowledge has constrained human cultural evolution [1]. Technology has had a varying impact on human cultural activities. On the one hand, it can democratize behavior. For instance, advancements in digital cameras and editing software have enabled individuals with limited technical knowledge to capture high-quality photographs and edit them effectively. On the other hand, the advent of new technologies can lead to increasingly specialized roles for humans. For instance, visual effects technology has led to the emergence of entirely new professions in film production, contributing to a cumulative increase in crew size [76]. Lastly, technological advancements can also lead to a reduction in human engagement in certain activities. For example, the advent of farming decreased the necessity for hunting, as people could grow food instead. It is yet to be seen whether the proliferation of increasingly intelligent machines will follow a similar trajectory. Will it democratize and increase human participation in creative fields, will it lead to new professions and increasing specialization, or could it lead to a decline in human-led creative expression, as we become increasingly reliant on intelligent machines [41]? Undoubtedly, human cultural exploration, in science as in arts, will increasingly be in collaboration with machines [77,78].

*Transmission*

The transmission of cultural information occurs via social learning, defined as learning that is influenced by the observation of another individual and/or its products [57]. This is in contrast to individual learning, where information is acquired, for instance, through trial and error. In humans, social learning tends to be imperfect, which means that transmission events are associated with a risk of losing cultural information. For example, if individuals observe a behavior incorrectly or do not fully understand it, they may not pass on the correct information to others. Over time, this can lead to the deterioration or even loss of cultural practices, which can have lasting consequences for a population.

Intelligent machines will increasingly be involved in the preservation and transmission of cultural information. Cultural evolution has supplied humans with increasingly efficient tools to preserve cultural information. The invention of writing, for instance, allowed humans to mitigate cultural loss by recording information in a more permanent way. Theoretical and empirical studies of cultural evolution have shown that the stability of cultural information strongly depends on both the size of the population that shares information [72] and the type of social learning mechanisms involved (with lower fidelity when transmission relies on mere observation than when it relies on verbal teaching, for instance) [79]. Besides serving as a persistent medium of cultural storage analogous to a book, machines can learn to seek and transmit information [78,80] and can act as conversational and pedagogical agents, similar to teachers [81]. This dual role has potential for drastically boosting cultural preservation by reducing cultural drift. For instance, LLMs have been harnessed to resurrect historical figures [82] and to revive languages teetering on the edge of extinction [83]. By fostering interactive cultural experiences for learners, these technologies can enhance the comprehension and retention of cultural information [84].

As machines store and transmit cultural information, they may reproduce and transmit biases inherent to human culture, for example through biased training datasets; but they also offer potential to mitigate those. Machine learning models reproduce various content biases inherent to their training data, including gender bias, racial or ethnic bias, bias for negative and threat-related information, and socioeconomic bias [68,85–89]. For instance, many LLMs make implicit assumptions about the gender associated with professions like nurse and doctor, reflecting the demographic perspective dominant in the training data [90]. When an LLM is used for revision and enhancement of human-written text, it can transmit and/or reinforce these biases [68]. Numerous strategies exist to address bias and fairness in machine learning models [91]. For instance, the demographic representation in the outputs of LLMs can be improved by using prompts featuring personas from a broad demographic spectrum [92]. However, LLMs present another challenge: they struggle to accurately represent languages and communities for which there is limited training data [68,93]. This limitation implies that LLM-based copy-editing could inadvertently lead to cultural erosion within these underrepresented communities due to losses in text reproduction. Nevertheless, LLMs, with their capacity to handle vast quantities of training data, can contribute to the preservation and even enhancement of cultural diversity, provided that the training data is carefully curated with attention to diversity and representation.

Cultural transmission can be impacted not only by the replication of biases but also by disparities between human and algorithmic biases. Examples of biases found in humans are confirmation bias [94], availability bias [95], and a bias towards specific symmetries [96]. Despite their reputation for undermining optimal decision-making, biases can actually reflect optimal decisions within a particular socio-environmental context under cognitive constraints such as memory, computation, or experience [97–100]. Similarly, machines may seize biases -- for instance, deep learning architectures assume spatial symmetries to improve training efficiency [94]. Importantly, humans and machines operate under different cognitive constraints [56] and inhabit distinct socio-environmental contexts. This may give rise to idiosyncratic biases. For instance, due to their enhanced computational capabilities, machines may display more utilitarian rationality [90]. Remarkably, humans often anticipate more utilitarian behavior from them [102] and might indeed, as we will elaborate in the following section, enhance this tendency. Irrespective of their origin, the effects of biases on culture tend to strengthen through repeated transmission [98,103,104]. This can restrict the spectrum of solutions that a population can derive, and even highly effective solutions may not be sustained if they conflict with pre-existing biases [105]. As such, the increased cognitive diversity—for instance with regard to biases—within human-machine societies has the potential to expand the collective capacity to maintain diverse cultural artifacts. Artifacts that might not be maintained by humans could potentially be maintained by machines. Conversely, in the transmission between humans and machines, a misalignment of biases can increase the risk of information loss [106].

Machines' increased computational capacity might additionally affect the feasibility of accumulating uncompressed information via a "big data" approach. The compressibility of information is the inverse of its algorithmic complexity: Compressing information is achieved by creating a rule that is shorter than a complete list of the data itself [107]. Compression is a key feature of both human cognition and machine learning [108]; any information-processing system must address a general trade-off between a truthful representation of the "raw" information and constraints on computation. The extent and nature of these constraints, however, differ between humans and machines [56]. Compression processes have an important role in human transmission to mitigate cultural loss: Raw information oftentimes is not learnable because it is too complex, whereas a compressed rule might very well be. Human language, for example, retains its expressivity by becoming learnable via the evolution of compressed structure [109]. Similarly, scientists develop, transmit, and revise theories as compressed representations of knowledge. Machine learning has the potential to reduce some of the constraints resulting from human computation, as the amount of data that can be processed is vastly increased. Consequently, information might be increasingly transmitted with low compression – in the form of "big data" – when predictive power is of ultimate importance and for some applications, the necessity to derive and transmit highly compressed rules or theories might be reduced [110,111]. For instance, with the advent of AlphaFold, scientists might focus on collecting and preserving further ground truth data on molecule structures to refine future models rather than refining and transmitting theories of atomic interactions. While symbolic representations may remain crucial for efficient computation, algorithmic-assisted discovery could lessen the need for their transmission, as these representations could be easily regenerated [112]. However, theories may retain importance in shaping human understanding and intuition, serving as essential tools for conceptualizing knowledge—a function that the framework we present in this work aims to fulfill.

*Selection*

Culture evolves in part through the selective retention of cultural traits. In the context of machine culture, selection can occur at a level where machines select what and from whom humans learn, at a level where humans select machine behavior, and at a level where there is selection between humans and machines.

When it comes to what humans learn, social learning strategies shape what, when, and whom we copy. These strategies can be broadly categorized into content- and context-based strategies [113,114]: content-based strategies consider what is learned, favoring—for instance—social over nonsocial information [115]. Conversely, context-based strategies attend to situational features, focusing on properties of a cultural model (e.g., their competence, success, prestige, knowledge, similarity), frequencies (e.g., most common behavior, rare behavior), or internal states of the learner (e.g., uncertainty, the cost of individual learning). Machines that help humans to navigate vast information spaces by (pre-)selecting cultural traits often reflect such social learning strategies.

For instance, content-based filtering algorithms aim to maximize the similarity between items a user previously showed interest in and unobserved items. Emulating context-based strategies in selection, ranking algorithms typically sort items according to a relevance score, which is based on the items' popularity [116,117]. Concurrently, collaborative filtering (CF) algorithms detect hidden patterns between items and users, achieving recommendations about novel items to similar users without using additional exogenous information about individual items or users [118] (see Fig. 4B).

Another dimension along which algorithms may influence the selective retention of cultural traits pertains to social networks, which form the backbone of information exchange. In this context, social ties are rewired as users follow algorithmic recommendations based on user attributes, such as popularity, or similarities in user preferences both in personal (X: "who to follow") and professional domains (LinkedIn "People you might know"). Link recommendation algorithms have the potential to shape the overall evolution of social networks [119–121]. X's "who to follow" recommendation was observed to disproportionately benefit those users who were already the most popular, fueling "the rich get richer" dynamics [121]. A growing body of research documents a complex but persistent and critical relationship between social networks' structure and collectives' ability to collaborate, coordinate, and solve problems [122–124] that ultimately shape cultural repertoires [73,125].

While machines have most commonly relied on exploiting explicit user preferences and historical behavior (e.g., ratings, engagement), there has been a growing interest in considering users' internal states to improve algorithmic recommendations. For instance, users might be uncertain about their preferences -- especially in domains in which they lack expertise. Bayesian approaches can model users' uncertainty and be used to update algorithmic recommendations as the user interacts with the system [126]. As another example, recommender systems might account for the cognitive cost of exploring items or learning about them by prioritizing items that are easier for users to evaluate or learn [127]. Affective recommender systems [128] use techniques such as natural language processing to make inferences about users' emotional states, and even combine them with other context information such as the recommendation domain (e.g., music or movies) [129–131]. While recommender systems have so far been mostly shaping user preferences implicitly (e.g., by optimizing the position or ranking of content), LLMs may accelerate developments where users are increasingly persuaded explicitly through interactive argumentation.

Downstream consequences of selection by machines may often be specific to particular environments, algorithmic models, and feedback loops. However, one feature generalizing across various contexts is that algorithms–by design of underlying business models–are often geared towards maximizing user engagement for profit [132]. In social networks, this may be achieved by promoting content congruent with users' past engagement or ingroup attitudes [133], or content that humans inherently attend to, such as emotionally and morally charged content [134,135]. One example for this is information that relates to threat or elicits disgust, as shown in transmission chain experiments inspired from cultural evolutionary theory [136]. The algorithmic amplification of such content may then feed back into human social learning, for instance inflating beliefs about the normative value of expressing moral outrage [137,138], increasing out-group animosity [139], or by creating echo chambers and filter bubbles [140–142]. It is important to note that user engagement is a signal of value to both users and platforms deploying algorithms, connecting them in complex feedback loops [143,144]: machines such as recommender systems react to user engagement, selecting types of content people engage with. Users also react to recommender systems, both directly in terms of clicking, viewing, purchasing but also in terms of what they produce, as content creators anticipate what will receive the widest distribution. These feedback loops, but also deliberate product design choices along with policy approaches, provide potent leverage points for aligning recommender systems with human values [145]. A promising approach to addressing this challenge could involve considering potential misalignments between users' engagement and their own preferences and identifying the boundary conditions determining when maximizing user engagement enhances user welfare and when it produces the opposite effect [146]. Algorithmic systems more generally offer powerful ways to bridge social divides, for instance by designing selection policies that steer users' attention to content that increases mutual understanding and trust [147,148], or by identifying and promoting links in social networks that can effectively mitigate polarizing dynamics [149,150]. Machine selection can also be deliberately geared towards fostering content diversity [151], or towards maximizing agreement among humans with diverse preferences [152].

Human preferences, in turn, can directly shape machine behavior, in particular through reinforcement learning with human feedback (RLHF) [153,154]. One way to conceptualize this is by viewing machines as students that generate arrays of solutions, with humans acting as teachers who select the most suitable ones. This process can nudge machines towards desirable properties like helpfulness, honesty, and harmlessness [154]. However, harvesting human annotators' preferences at scale through RLHF can induce machines to adopt behaviors unintended by the deploying organizations, exemplified by chatbots that increasingly endorse inflated political views or express a heightened desire to avoid shutdown [155]. Human influence on machine behavior also occurs through more subtle pathways: through training on human text alone LLMs picked up a tendency to repeat users' stated views [155]. Machines' attention to human feedback may create selection pressures towards pleasing human interlocutors. For instance, humans may favor machines catering to non-factual stories and narratives that match concepts and ideas preferred by human cognition [156], such as those appealing to intuitive expectations about the natural world [157,158], or to specific religious practices [159].

Humans select machine behavior also through direct creation and curation of training data for machine learning, and through more indirect interactions with machine-generated outputs. Despite efforts to watermark machine-created content [160], machine and human-made content will increasingly intertwine [161]. For instance, it is estimated that many supposedly human-written texts on crowdsourcing platforms are already augmented by machines [162]. Consequently, it seems inevitable that future machines will be trained on mixed human-machine content, forming part of a larger feedback cycle between content generation and selection. The human element in this cycle may prove crucial, for instance in preventing 'model collapse' -- a dynamic where repeated training on machine-generated data narrows its outputs to very few traits [161]. This phenomenon stems from a classification bias that favors more prevalent classes [163], a bias that is amplified through successive iterations of learning [164]. By contrast, when encountering similar challenges, human culture might preserve diversity by utilizing biases with counteracting effects, such as endorsing local conformity [165]. That said, it seems likely that machines could recover similar strategies to maintain cultural diversity without human involvement.

Yet another pathway for human selection on machine behavior pertains to general machine properties. For instance, humans may choose between different intelligent machines available on the market based on factors such as preferences, cost, usefulness, harmfulness, and alignment with regulatory requirements. As such, the language model LLaMA recently gained attention for its relatively smaller size, making it more cost-effective to use [166]. Conversely, ChatGPT outperformed many comparable models due to its superior accessibility and helpfulness [154]. Human demands towards machines are likely to change over time, shaping the trajectory of machine culture similar to other historic cases where technologies evolved in response to changing human demands (e.g., the wheel from wooden wheels on carriages to rubber tires on automobiles). Crucially, if intelligent machines are designed and evaluated by non-representative experts, these systems run the risk of unintentionally reproducing and intensifying the biases inherent to their selectors [167].

Selection is also bound to occur at a level where machines or humans are favored over one another. For instance, the ability of machines to process vast amounts of information both quickly and accurately affords them a competitive edge in numerous cognitive tasks, such as strategic gameplay and information retrieval. Relatedly, due to their cost-effectiveness and efficiency, intelligent machines may grow into the main workforce across various professional domains [38,168]. However, analogous to how the invention of the car did not diminish interest in running as a sport, the proliferation of machines may not curtail human interest in intellectual pursuits; instead, it might redirect the focus from necessity to leisure and entertainment. Meanwhile, across various contexts, humans might favor other humans over machines due to their shared experience. For instance, even though present-day machines can conceive messages perceived as more empathic than messages conceived by humans [169,170], such messages may be perceived differently once recipients become aware of their artificial origin. This phenomenon, which could be referred to as the "artificial empathy paradox" [171] may, at least in part, arise from the very fact that empathizing is effortful for humans [172] and that, as such, it conveys a motivational social signal to others that is made void by machine-involvement. Selection between humans and machines does not imply that one agent dominates over the other in any cultural niche; oftentimes, one will augment -- rather than fully replace -- the other; other times, the presence of the other may trigger the development of new skills, or roles.

# Grand challenges and open questions

We now suggest a broad research agenda for computational and behavioral scientists interested in the phenomenon of Machine Culture.

*Measurement*

One major open challenge is to quantify how much of human cultural dynamics can be attributable to algorithmic processes. For instance, it is difficult to completely disentangle the effects of ranking and recommendation algorithms on culture from alternative processes of human social learning, such as communication technology, institutions, and social practices. Since the inception of human culture, derived tools have had an important role in shaping cultural processes, making the establishment of a baseline a challenging question in itself. Getting good estimates is a precondition to optimizing for the usefulness of these algorithms while avoiding undesirable cultural impact such as polarizations [173]. This challenge is reflected in research on "filter bubbles", which has moved from considering algorithmic curation as decisive force shaping online engagement[141] towards acknowledging the influence of users' own choices on social media ranking algorithms [174] and search engines [175]; thus highlighting the intricate feedback loops between machine and human decisions. While we hope that it will be possible to find appropriate metrics for the cultural impact of machines, the intricacy of this problem qualifies it as an open grand challenge.

A complementary question is how to quantify the influence of machine-generated artifacts – e.g., artwork, literature, music – on human cultural production in these areas. As generative AI becomes more commonplace, distinguishing intrinsic human culture from machine-generated culture or machine-influenced culture becomes even more challenging, especially as watermarking techniques may not be universally adopted. Despite popular media claims about machine-generated art developing its own unique style [176], we do not yet have a reliable way of verifying these claims, let alone assessing machine-to-human cultural transmission [106].

Another measurement concern relates to the quantification of the cultural regularities encoded in LLMs and other AI models. Even prior to the rise of LLMs, social media platforms like Facebook already possessed fairly detailed quantitative models of cultural regularities and differences [177], developed mainly for the purpose of marketing to particular demographics. However, as LLMs are trained from curated datasets, and fine-tuning using human feedback, quantifying the biases that are introduced and/or mitigated by these models is crucial [178,179].

*Societal Decision-making*

We currently observe a rapid increase in the diversity of AI models, including LLMs, accelerated by the open-source software movement. However, market forces, such as regulation and market power, may result in a world dominated by a small number of monolithic models. This raises the possibility of reduction in cultural diversity, as major social, political, and economic forces try to shape global machine culture to match their preferences. This process may be amplified by feedback loops, in which LLMs train on an ongoing basis from synthetic data, or from human data that contains much machine-generated text. Preliminary evidence points to the possibility of model collapse, with the models losing diversity and converging to a state with low variance [161].

Conversely, we face a potential 'Tower of Babel' scenario. As AI models become increasingly personalized, conforming to and reinforcing our individual worldviews, they risk engendering an unprecedented fragmentation of our shared perception of the world. In the biblical story (Genesis 11:1-9), the construction of the tower led to a divine intervention that scattered humanity and confounded languages. Drawing a parallel, if we continually interact with machines that echo and affirm our preconceived notions, we risk isolating ourselves within ideologically and culturally homogenous echo chambers. Such fragmentation can stifle meaningful dialogue, breed misunderstanding, and, ultimately, fracture our shared future vision.

Against this background, a key research agenda is to quantify the degree to which a given AI model, or ecosystem of models, exhibits uniformity or diversity. It is also imperative to understand what constitutes a 'healthy' level of diversity, that retains local sovereignty while also fostering collective human flourishing.

Ensuring that AI models, such as LLMs, reflect the beliefs and values of a given community requires mechanisms for societal decision-making about what knowledge goes into the models [180]. Furthermore, humans exhibit variation in their ethical expectations towards machines, both within and across cultures [181]. This raises questions about how to best aggregate diverse, potentially conflicting preferences to arrive at an agreeable outcome [182]. A number of interesting ideas are emerging, from voting on different algorithmic policy-makers [183] to utilizing LLMs to summarize diverse human opinions [184] and generate consensus statements [152].

A related challenge is how to ensure long-term monitoring of Machine Culture. Similar to the notion of human-in-the-loop control of intelligent machines, we can aspire towards society-in-the-loop control of the complex phenomena of Machine Culture [185].

Suppose a community knows which cultural beliefs and values it wants to encode in an AI model. The next question is how to ensure that these are indeed present in the model. One approach is to carefully curate the dataset on which the model is trained [186,187]. Another, increasingly used approach is fine-tuning based on RLHF [188]. Yet another approach is to fine-tune using constitutional rules provided by humans [189]. The relative merits and drawbacks of these various approaches are still not well understood. There is a need for methods to check which cultural beliefs and values have been learned, inferred, or encoded in a given AI model and the degree to which it aligns with a target culture.

*Long-term Dynamics & Optimization*

In all likelihood, the future of culture will be hybrid, with cultural artifacts–scientific theories, industrial processes, art, literature–being created by a combination of human and machine intelligence. This raises a suite of open questions relating to the long-term dynamics of human-machine co-evolution. These dynamics may lead to diverse phenomena ranging from different forms of human-machine mutualism, to Red Queen effects (an evolutionary arms race between humans and machines) that characterize the co-evolution of both forms of intelligence [190].

A related question is how to optimize the aforementioned dynamics, in order to combine human and machine intelligence in an ideal or safe manner [75]. This may be relevant to questions of risk mitigation. Although experts disagree on the timescales and the degree of risk involved, the potential of superhuman Artificial General Intelligence (AGI) poses a possible existential threat to the human species [191]. Cultural evolution provides a useful framework for navigating this challenge. Specifically, cultural evolution processes take place today at multiple scales, with human collectives–e.g., companies, universities, institutions, cities, nation-states–acting as the units of selection [192]. This multi-level selection can, in principle, operate at the level of human organizations augmented by intelligent machines and (eventually) superhuman AGI. Engineering this evolutionary process can provide means for ensuring human survival and agency in the long run.

# Conclusion

We asked GPT-4 to first write a compressed version of this Perspective and then to provide a conclusion. It suggested the following (minimal editing to align nomenclature has been applied). The symbiosis of human and machine intelligence is forging a new epoch of cultural evolution. This Perspective highlights the transformative role of intelligent machines in reshaping creativity, redefining skill value, and altering human interactions. Central to the discourse is the triad of cultural evolution: variation, transmission, and selection, and how machines interface with each. The interaction is multifaceted, from generative AI birthing novel cultural artifacts to recommendation algorithms influencing individual perspectives. However, the crux remains in understanding and navigating the challenges and opportunities that arise from this hybridization of culture. As the imprints of intelligent machines grow deeper, it's imperative to ensure a harmonious co-creation of culture where both human and machine augment, rather than eclipse, each other. This not only broadens the horizons of cultural exploration but also fortifies the tapestry of human experience in the age of intelligent machines.

# Author contributions

## Acknowledgments

LB thanks Michiel Bakker for useful discussions. The authors thank Bramantyo Supriyatno for supporting the formatting of the manuscript. The conclusion was created by GPT-4 based on a summary (also created by GPT-4) of this manuscript with minimal editing to align nomenclature. JFB and MD acknowledge IAST funding from the French National Research Agency (ANR) under grant no. ANR-17-EURE-0010 (Investissements d'Avenir programme)

## Competing interests

The authors declare no competing interests.